# Electrochemical doping in H-terminated diamond films: Impact of O-functionalization and insights from in-situ Raman spectro-electrochemistry

N. Mohasin Sulthana [a,b], P.K. Ajikumar [b], K. Ganesan [a,b,1]

[a] *Homi Bhabha National Institute, Training School Complex, Anushakti nagar, Mumbai-400094, India*

[b] *Materials Science Group, Indira Gandhi Centre for Atomic Research, Kalpakkam- 603102, India*

**Abstract**

The p-type surface conductivity of H-terminated diamond (HD, H-diamond) has created new path ways for developing diamond based electronic devices as well as chemical and bio-sensors. However, the hydrophobic nature of the HD surface can negatively impact device performance due to its low wettability. Herein, we report the study on polymer electrolyte-gated field effect transistors (EGFETs) fabricated using pristine and partially O-terminated HD films. The HD surface is transformed from hydrophobic to moderate hydrophilic by partial O-termination. Also, the sheet resistance of the HD surface increases from 7.6 to 18.7 kΩ/□ while the sheet hole density decreases from 10.5 to 4.8 x $10^{12}$ cm$^{-2}$ upon partial O-termination. Consequently, the ON/OFF ratio of the EGFET devices decreases from ~ 40 to 14 and the maximum transconductance declines from of -150 to -7.9 μS/V, but the areal capacitance increases from 7.8 ± 3.6 to 27.1 ±10.1 μF/cm$^2$ with partial ozonation on HD surface. In addition, the in-situ Raman measurements in HD EGFET provide direct experimental evidence of a gating-induced blue shift and linewidth broadening of the diamond Raman band, which are associated with strong electron–phonon coupling. This work highlights the significant impact of the partial O-termination on the performance of the HD EGFET devices and effect of electrochemical gating on the phonon behaviour of the H-diamond.

---

[1] Corresponding author. Email : kganesan@igcar.gov.in ( K. Ganesan)

# 1. Introduction

Diamond exhibits a lot of attractive properties including chemical inertness, electrochemical stability, intrinsic low noise, large electrochemical potential window, and biocompatibility. These characteristics make it an exceptionally promising candidate for a wide range of applications such as chemical and bio-sensors, bioelectronic and electrochemical devices [1–4]. Also, over the past two decades, the discovery of surface conductivity (SC) arising from surface functionalization has opened new opportunities for diamond based electronic devices [5–7]. Here, the interfacial properties play a crucial role to evince the successful platform for such applications. However, the interfacial properties of the diamond surface are strongly dictated by its surface functionalization. For instance, the diamond surface can exhibit either hydrophobic or hydrophilic behaviour depending upon its surface functionalization. Hydrogen and fluorine functionalization leads to hydrophobicity, while oxygen and nitrogen functionalization leads to hydrophilicity of the diamond surface [8,9]. Hence, the effect of these surface functionalization cannot be ignored when considering the aqueous electrolyte-based devices.

Among the various types of surface functionalization, the hydrogen (H) and oxygen (O) termination on diamond are the most extensively studied due to their intriguing surface electronic characteristics. Particularly, H-terminated diamond (HD, H-diamond) surface possesses negative electron affinity (NEA) which leads to high SC facilitated by upward band bending through surface transfer doping [10–12]. Further, HD surfaces are strongly hydrophobic and exhibit a stable surface. In contrast, O-terminated diamond (OD) surfaces exhibit positive electron affinity, resulting in an insulating surface with downward band bending [12,13]. Further, OD surfaces display a strong hydrophilicity with active catalytic nature. These distinct characteristics highlight the contrasting surface electronic properties of HD and OD films.

The hydrophobicity plays a major role in limiting the interaction of water on the surface leading to depletion of water molecules in the near surface area [14]. However, the impact of this surface wetting property has been barely considered to explain the operation of any devices. In biosensor and bioelectronic devices, the change in electrochemical potential happens across the interface wherein water molecules modulate the concentration of charge carriers. As a consequence, its conductivity changes that leads to the sensor signals. Additionally, the conductivity variation due to potential-dependent charge carriers significantly influences the interface capacitance which is a key factor in the performance of electrolyte mediated field effect transistors (FET). The depletion of water near the surface further impacts the interface capacitance by altering the local dielectric constant [14]. To study such an effect of hydrophobicity on the diamond surface, the concept of FET with electrolyte gating can be conveniently used. Notably, the gating with solid electrolyte offers advantage of achieving high carrier concentration at low gate voltages, without requiring an insulating dielectric material. The charge separation in the electric double layer formed between gate and conducting channel is very small that results in high interfacial capacitance. Studies on the electrochemical doping of H-terminated diamond are still very limited, and those involving partial oxygen termination are even scarcer.

Raman spectroscopy, however, serves as a powerful, successful, and non-destructive tool for probing changes in lattice dynamics induced by doping. For two-dimensional systems such as graphene, $MoS_2$, and phosphorene, Raman spectro-electrochemistry has been used to estimate the number of layers and doping concentration, as well as to characterize the effect of doping on electron-phonon coupling strength and associated phase transitions [15–18]. In these 2D systems, electrochemical doping leads to phonon renormalization, including frequency shifts, variations in phonon lifetime, and the breakdown of selection rules, which provides valuable insights into doping effects. While the effects of conventional dopants such as boron and nitrogen on diamond's phonon properties have been well investigated using Raman spectroscopy as a function of doping concentration, the impact of electrochemical doping leading to high hole density in the 2DHG on diamond's phonon behaviour remains unexplored, making it a subject of growing interest [19–21].

In this study, we endeavour to fabricate an easy-to-process electrolyte-gated field effect transistor (EGFET) using polyethylene oxide (PEO) and $LiClO_4$ polymer electrolyte on surface conducting polycrystalline diamond films. To explore the impact of surface oxygen on the transistor characteristics such as ON/OFF ratio, transconductance and threshold voltage, EGFET based on both pristine and partially O-terminated HD were analysed. The impedance spectroscopy was also employed to investigate the interface between PEO electrolyte and functionalized diamond surface. In addition, the in-situ Raman spectroscopy was carried out to investigate the effect of electrochemical gating on the phonon dynamics of the HD sample.

## 2. Experimental details

Polycrystalline diamond films were deposited on $SiO_2$/Si substrate by hot filament chemical vapour deposition using feed-stock gases $CH_4$ and $H_2$ in the ratio of 2:200, as per the details given elsewhere [8]. After growth, hydrogen termination was achieved on diamond surface by admitting ultra-high pure hydrogen into the chamber at a flow rate of 200 sccm for 20 min under working pressure of ~ 40 mbar. The substrate temperature was maintained at 800 °C during H termination. To partially O-terminate the HD surface, the HD film was exposed to ozone atmosphere for duration of 30 s using UV/ozone pro cleaner (UV ozone cleaner, Ossila) at room temperature under ambient pressure. Later, both HD and partially O-terminated HD films were exposed to air atmosphere for 2 days to reach surface charge transfer equilibrium state. The selection of a 30 s ozonation duration was based on detailed surface characterization of H-diamond, including electrical conductivity measurements and X-ray spectroscopy performed after ozonation for varying exposure times [8].

The surface morphology of the functionalized diamond films is analysed using field emission scanning electron microscopy (FESEM, Carl-Zeiss, Switzerland). The wetting characteristics of the surface functionalized diamond films were analysed by water wetting contact angle (WCA) measurement. The structural properties of the films were studied using Raman spectroscopy (In-via, Renishaw, UK) with a diode laser having wavelength of 532 nm and a grating of 1800 grooves/mm in backscattering geometry with 50x objective lens. Source and drain contacts were prepared by depositing Pd (20 nm) with an over layer coating of Ag (100 nm) on the samples by shadow masking method using thermal evaporation. The Ohmic nature of the contacts were confirmed through I-V characteristics. The separation between source and drain is ~ 500 μm. The sheet resistance, carrier density and mobility of the pristine and partially ozonated HDs films were measured using Hall measurement on three independent samples.

Top gating was attained using solid polymer electrolyte consisting of $LiClO_4$ and PEO (Mw = 100,000 sigma Aldrich). First, PEO and $LiClO_4$ (1 and 0.3 g, respectively) were mixed with 30 ml anhydrous methanol, as reported for $TaS_2$ thin flakes [22]. Then, the mixture was stirred for 8 hrs maintaining the temperature at 50 ºC. Afterwards, the mixture was drop-casted on the surface functionalized diamond in between the source and drain electrodes. A sharp platinum needle was used as gate electrode and the schematic of the EGFET device is shown in Fig. 1. The PEO electrolyte/diamond interface was characterized with electrochemical impedance spectroscopy (EIS). EIS measurements were performed over a frequency range of 0.1 - 1000 Hz by applying a 10 mV sinusoidal voltage between the gate electrode and the source terminal, using Metrohm-Autolab electrical work station (model PGSTAT302N, Netherlands) on two independent diamond EGFET devices. The impedance data were analysed using EIS Spectra Analyzer software with an appropriate equivalent circuit model. EGFET characteristics were measured using Agilent B2902A precision Source/measure unit. Furthermore, the in-situ Raman spectro-electrochemical measurements were performed using a micro-Raman spectrometer in reflection geometry, equipped with a 2400 grooves/mm grating and a 100× objective lens. The Raman spectro-electrochemical measurements were repeated at least three times at the same spatial location for each gate voltage on the EGFET device. The obtained spectra were fitted using a Lorentzian profile, and the resulting fitting parameters were analysed statistically.

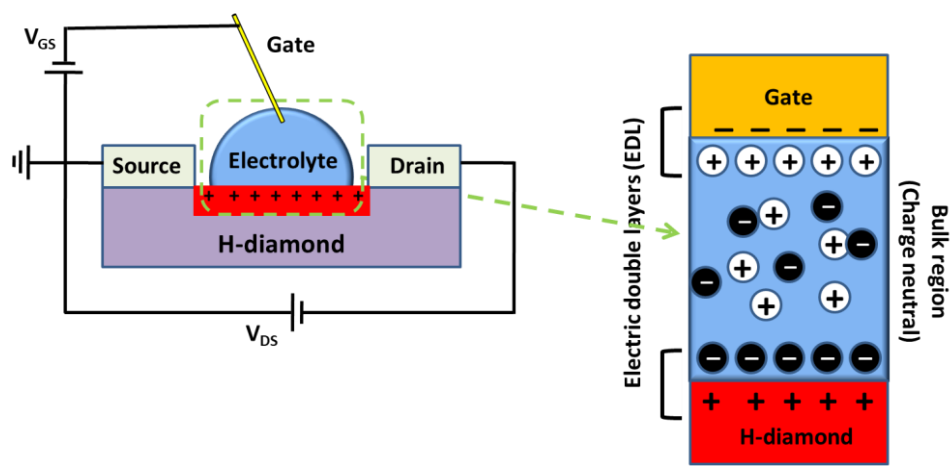

Fig. 1. Schematic of H-diamond based EGFET device

## 3. Results and discussion

### 3.1 Surface morphology and Raman spectroscopy

The FESEM surface morphology of the as grown HD film shows the multi-faceted structure with grain size of ~ 1 μm, as shown in Fig. 2a. The surface morphology predominantly exhibits the rectangle and triangle faceted structures corresponding to (100) and (111) planes, respectively. The thickness of the film is about 2.5 μm. Fig. 2b displays the Raman spectrum of the as grown HD film. The sharp Raman band at 1332 cm$^{-1}$ corresponds to the zone centre phonon with $T_{2g}$ symmetry of diamond [21]. The FWHM of the Raman band is ~ 7.6 cm$^{-1}$ indicating the high structural quality of the microcrystalline diamond film. In addition, the slight upward slope in the Raman spectrum indicates the presence of non-diamond carbon species, such as sp²/sp¹ hybridized bonding, primarily located at grain boundaries, along with contributions from impurity-related electronic states, potentially including nitrogen. Although nitrogen was not intentionally introduced, its incorporation during CVD growth is difficult to entirely prevent.

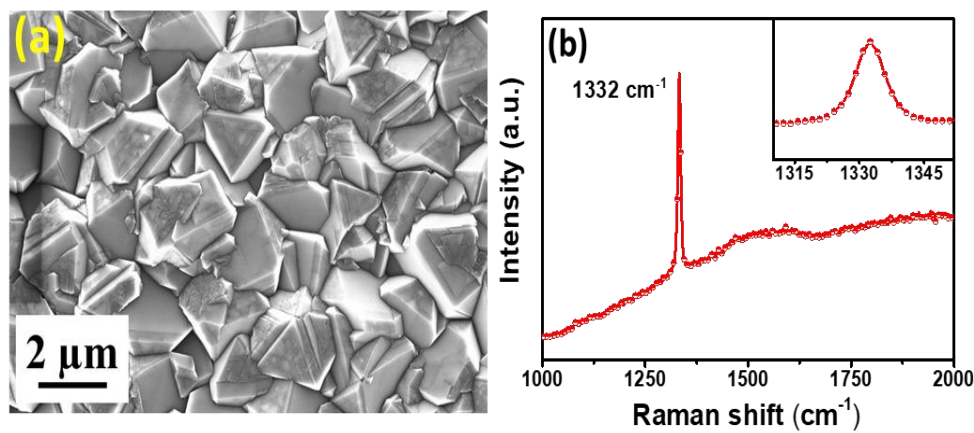

Fig. 2. (a) FESEM micrograph and (b) Raman spectrum of the as grown HD film. The inset shows the magnified part of Raman band around 1332 cm$^{-1}$

### 3.2. Wetting contact angle analysis

Figure 3 shows the change in WCA of HD surface after partial ozonation for 30 s. The measured WCA of 99° of HD surface confirms its intrinsic hydrophobic nature due to the dominating van der Waals interaction which weakens the water physisorption on the surface [9]. However, this hydrophobicity evolves into hydrophilicity, when a small fraction of oxygen functional groups is incorporated at H-site on the surface. The surface oxygen functional groups interact with the polar medium (water) and leads to reduced water WCA [23]. Hence, the WCA is found to be reduced down to 74° for partially O-terminated HD surface, that confirms the partial substitution of oxygen functional groups at H-sites on the surface. As demonstrated in our earlier X-ray photoelectron spectroscopy study on H-diamond, the O/C ratio increases systematically with ozonation duration; for instance, it rises from 1.3 to 2.0 % after 30 s of ozonation. The XPS results further indicate that the partially O-terminated H-diamond surface is predominantly functionalized with hydroxyl (C–OH) and epoxy (C–O–C) groups. In addition, the progressive incorporation of oxygen-containing functionalities is supported by WCA measurements [8].

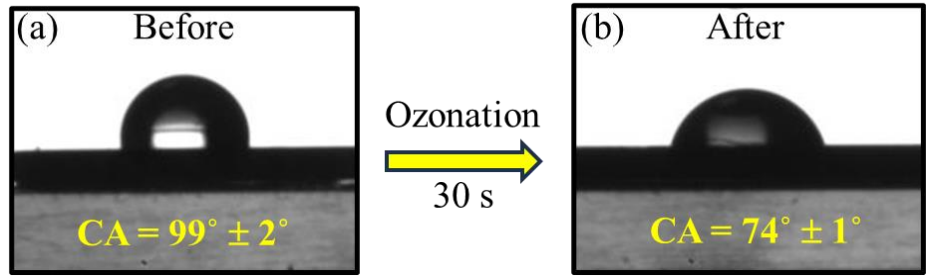

Fig. 3. Wetting contact angle of (a) pristine HD and (b) partially O-terminated HD

**3.3 Surface electrical characteristics**

Figure 4 represents the I-V characteristics of the pristine and partially O-terminated HD samples. The linear I-V characteristics affirm the Ohmic behaviour of the metal electrodes on the films. As depicted in the table 1, the sheet resistance of the pristine HD sample is 7.6 k$\Omega$/□ with hole density of $1.1 \times 10^{13}$ cm$^{-2}$ and mobility of 77.0 cm$^2$/Vs. However, the sheet resistance of the partially O-terminated HD sample is increased to 18.7 k$\Omega$/□ with hole density and mobility of $4.8 \times 10^{12}$ cm$^{-2}$ and 70.9 cm$^2$/Vs, respectively. Here, we provide a brief account on the surface transfer doping model that describes the SC of the HD. The NEA of the HD is the prerequisite for its SC. The chemical potential of atmospheric molecules is ~ 4.8 eV that lies below the valence band maximum of the diamond. When these atmospheric molecules are adsorbed on the HD surface, spontaneous electron transfer takes place from valence band maximum of the diamond to the lowest unoccupied molecular orbital of the adsorbed molecules [10]. It happens through electrochemical redox reactions that results in hole accumulation on the diamond surface. This hole accumulation creates an upward band bending at the surface and also, the surface Fermi level ($E_F$) moves below the valence band maximum ($E_v$) of the diamond. Hence, the high SC is created on HD film surface. Comparing to pristine HD sample, the partially O-terminated HD sample possess higher sheet resistance in consequence of lower hole density, due to the partial substitution of surface H- atoms by O-atoms during ozonation. The local EA changes from negative to positive at surface C-O site which lowers the preferable condition for charge transfer. Hence, the surface hole density is lowered in partially O-terminated HD surface comparing its pristine surface.

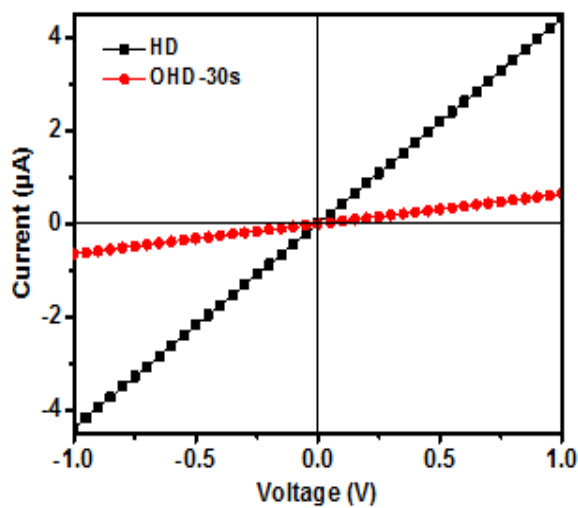

Fig. 4. Current versus voltage characteristics of pristine and partially O-terminated HD surfaces

Table 1. Electrical parameters of the pristine and partially O-terminated HD samples obtained from hall meausrement

| Sample | Sheet resistance (k$\Omega$/sq) | Hole density ($10^{12}$ cm$^{-2}$) | Hole mobility (cm$^2$/Vs) |
|---|---|---|---|
| HD | 7.6 ± 2.1 | 10.5 ± 3.6 | 77 ± 4 |
| OHD-30s | 18.7 ± 8.3 | 4.8 ± 3.2 | 70 ± 11 |

### 3.4. Electrochemical impedance spectroscopy

Figure 5 displays the variation of impedance and phase of PEO electrolyte/diamond interface in the frequency range from 0.1 to 1000 Hz. The experimentally obtained impedance data is fitted with an equivalent circuit having resistance and constant phase element in parallel representing the charge transfer resistance and capacitance respectively, along with one more resistance in series representing the resistance of the electrolyte. Table 2 provides the extracted parameters from the electrochemical impedance analysis. The resulting resistance of electrolyte is found to be ~ 5 x $10^4$ $\Omega$ for both HD and partially O-terminated HD samples. Further, the charge transfer resistance of 2.9 ± 1.3 and 5.1 ± 1.1 M$\Omega$ are obtained for pristine and partially O-terminated HDs, respectively. While the electrolyte resistance is same for both the samples, the charge transfer resistance of partially O-terminated HD is higher than that of pristine HD due to higher sheet resistance of partially O-terminated HD surface. The areal capacitance is estimated to be ~ 7.8 ± 3.6 and 27.1 ±10.1 µF/cm$^2$ for pristine and partially O-terminated HD surfaces, respectively. The areal capacitance of the partially O-terminated HD sample is higher than that of the pristine HD sample due to the higher wettability which helps to increase the electric double layer capacitance.

Note that the high SC of HD is due to the surface transfer doping mechanism wherein the water adlayer acts as electron sink that induces two-dimensional hole gas on the diamond surface. Hence, the hole accumulation on the HD surface is limited by hydrophobic gap with water adlayer in the vicinity of the diamond surface. In addition, the hydrophobic gap affects the interfacial properties of the PEO electrolyte/diamond interface and it leads to lowered areal capacitance. However, such a hydrophobic gap can be reduced by partial O-termination and also, such a partial O-terminated HD can retain the advantages in terms of high SC of HD.

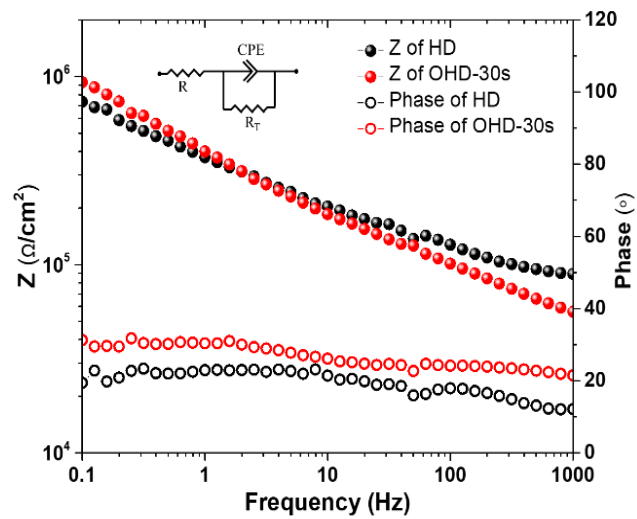

Fig. 5. The variation of impedance and phase of PEO electrolyte/diamond interface as a function of frequency

Table 2. The interface electrical properties of the pristine and partially O-terminated HD films obtained from electrochemical impedence analysis

| Samples | Electrolyte resistance (k$\Omega$) | Charge transfer resistance (M$\Omega$) | Areal capacitance ($\mu F/cm^2$) |
|---|---|---|---|
| HD | 50.1 ± 1.5 | 2.9 ± 1.3 | 7.8 ± 3.6 |
| OHD-30s | 50.1 ± 1.5 | 5.1 ± 1.1 | 27.1 ± 10.1 |

## 3.5 EGFET characteristics

Figures 6a and 6c depict the $I_{ds} - V_{ds}$ characteristics of the EGFET from pristine and partially O-terminated HD devices, respectively. Here, the variation of source – drain current ($I_{DS}$) is measured for applied drain-source voltage, $V_{DS}$, in the range from 0 to 0.2 V at different gate voltage ($V_G$) that varies from + 0.6 to – 3.6 V at a step size of 0.1 V. The Figs. 6a and 6c clearly indicate that the $I_{DS}$ can be controlled through $V_G$. Further, the Figs. 6b and 6d show the variation of surface transconductance ($I_{DS}/V_{DS}$) as a function of $V_G$ in the voltage range from + 0.6 to − 3.6 V for a fixed $V_{DS}$ at 0.1 V for pristine and partially O-terminated HD devices, respectively. These results suggest that the EGFETs can be stably operated without taking into consideration about the partial O-termination on the HD surface. The source-drain saturation current is about 40 and 14 times higher than that of cut-off region current for pristine and partially ozonated HD devices, respectively. These EGFET characteristics were attained reproducibly using PEO and LiClO4 electrolyte as a gate material.

We note here that the channel conductance of the EGFET device is controlled by electrostatic charge at the PEO electrolyte/diamond interface wherein no faradaic reaction takes place. Hence, the electrolyte as such do not affect the channel of the HD EGFET device since there is no charge transfer across the interface. The gate electrode controls the PEO electrolyte/diamond interfacial potential that actually the changes channel conductance by the electrolyte induced capacitive charging on the surface. When gate potential is increased in negative direction, the Fermi energy level of the diamond surface is pushed further below the valence band maximum of the diamond surface that leads to additional accumulation of holes on the diamond surface [24]. In contrast, when the gate potential is increased in the positive direction, the Fermi energy level of the diamond surface is pushed above the valence band maximum that leads to decrease in the already accumulated holes on the diamond surface. Hence, the increase in hole density on the diamond surface by applied gate potential leads to the increase in SC of the diamond.

Further, the threshold voltage of the EGFET devices is extracted using extrapolation method in the saturation region from the transfer characteristics [25]. The threshold voltage is found to be ~ -2.2 and -2.4 V for pristine and partially O-terminated HDs, respectively. In addition, just above their threshold voltage over a range of gate voltages, the drain - source conductance is linear with a slope of -150 and -7.9 µS/V for pristine and partially O-terminated HDs, respectively. Moreover, the hole mobility extracted in the linear region from the transconductance is ~ 192 and ~ 3 cm$^2$/Vs for pristine and partially O-terminated HD EGFETs, respectively. Thus, the transconductance and the hole mobility are decreased more than one order as a result of partial O-termination. Despite employing polycrystalline diamond films, the HD EGFETs in this study exhibit transconductance values comparable to those reported for fully H-terminated single-crystal diamond EGFETs by Denkerl et al. [24] (~100 µS/V), although the ON/OFF ratio and hole mobility are relatively lower. Moreover, the large discrepancy observed between the mobility of partially O-terminated H-diamond measured by Hall experiments and that extracted from EGFET I–V characteristics primarily arises from the different physical conditions under which these measurements are performed. The mobility obtained from EGFET measurements represents an effective mobility that is strongly influenced by interface quality, gate capacitance, surface scattering, and trap states. Consequently, it can deviate significantly from the intrinsic drift mobility determined through Hall measurements.

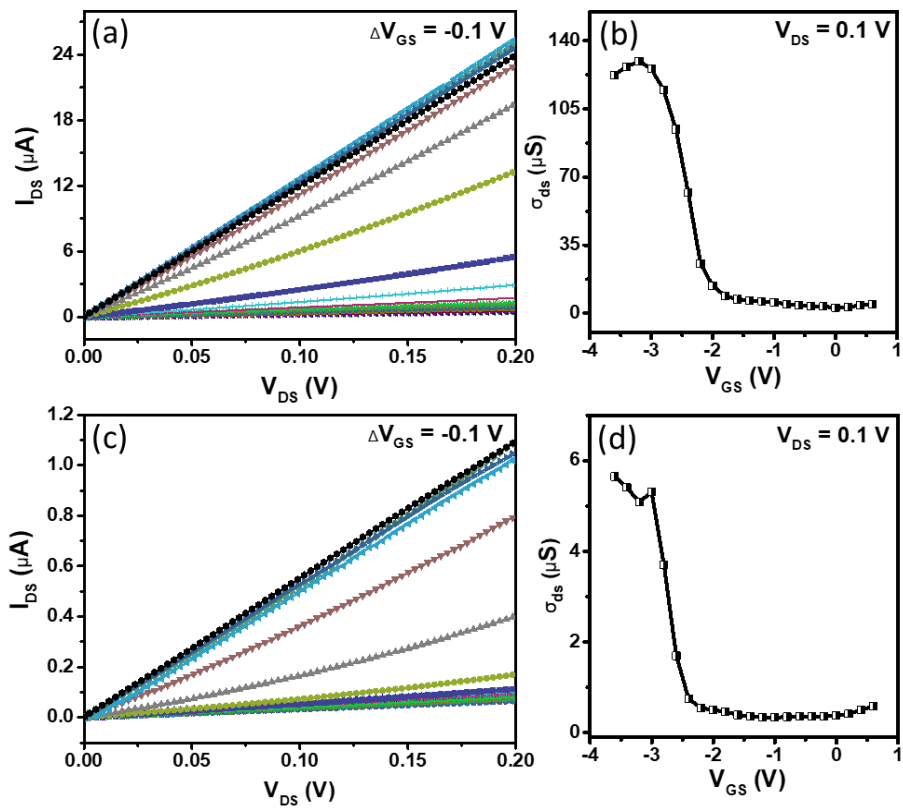

Fig. 6. (a and c) The current versus voltage between source and drain at different gate potentials and, (b and d) sheet conductance at $V_{DS}$ of 0.1 V versus gate potential for pristine and partially O-terminated HD EGFET devices, respectively.

Further, their transconductance shows their maximum at gate voltage of -2.4 V for pristine HD and -2.8 V for partially O-terminated HD. This is the voltage that shows the highest sensitivity of the drain-source current change towards the gate potential. Moreover, both the threshold and peak transconductance voltage are increased in the negative direction for the partial O-termination. The partial O-termination causes a decrease in upward surface band bending due to the reduction in surface carrier density. Consequently, the gate voltage applied for the partially O-terminated HD EGFET is inadequate to reach the same level of carrier accumulation as in the case of pristine HD EGFET. Thus, this resultant difference in inherent upward band bending between the pristine and partially O-terminated HDs causes this threshold voltage shift.

The comparison of the transconductance of the EGFETs is the effective approach to evaluate their performance. The higher transconductance of HD device reflects the apparent improvement of its performance over the partially O-terminated HD. Further, the enhanced transconductance in HD can be ascribed to the increased modulation of the hole concentration on the diamond surface that results in increased current modulation of HD. Although the areal capacitance of PEO electrolyte/diamond interface in partially O-terminated HD is higher, that can facilitate a higher surface hole density, the current modulation is lower than that of HD. The limitation for increasing hole density is mainly arises due to a significant amount of local positive electron affinity (PEA) in partially O terminated HD. The local PEA nature at the O-sites on the surface resists the charge transfer doping process and hence, the reduction in hole accumulation on the surface [8,26]. Further, the action of O-sites as scattering centres significantly reduces the carrier velocity, consequently the hole mobility decreases [27]. Thus, although the elevated areal capacitance could improve the performance of partially O-terminated HD EGFET, the increased charge transfer resistance and reduced hole mobility impact negatively on the overall performance. Nevertheless, the partial O-termination on HD surface did not deteriorate the FET device characteristics significantly. Since the high wettability is a prerequisite for high performance biosensors based on diamond films, partially O-terminated HD based devices have high potential for such applications. For example, Garrido et al. [28] reported that an H-terminated diamond ion-sensitive FET was largely insensitive to pH variations, whereas a surface with partial O-termination achieved through ozone treatment exhibited pronounced pH sensitivity. The capability to monitor pH changes further enables indirect detection of chloride concentrations in blood or sweat by measuring $Cl^-$ levels, which is relevant for cystic fibrosis diagnosis [29]. Similarly, Sakai et al. [30] demonstrated that partially O-terminated diamond solution-gated FETs operated stably, despite a negative shift in threshold voltage. Notably, these devices showed enhanced sensitivity to positive ions and neutral molecules, with minimal interference from negative ions. Thus, partial O-termination presents a clear trade-off: improved electrostatic coupling and enhanced sensitivity for chemical and biosensing applications, accompanied by some degradation in conventional FET characteristics such as hole mobility and ON/OFF ratio. For bio- and electrochemical sensing, however, high interfacial capacitance and improved surface wettability are often more critical than carrier mobility. Therefore, partially O-terminated H-diamond devices remain highly relevant for practical sensing applications, despite moderate reductions in standard FET performance metrics.

### 3.6 *in-situ* Raman spectro-electrochemistry

Figure 7(a) shows the Raman spectra of diamond measured as a function of gate voltage for the HD EGFET. Under unbiased conditions, a sharp first-order optical phonon peak appears at approximately 1332.4 cm$^{-1}$. Although no significant changes are visually apparent in the Raman peak with applied bias, fitting the data reveals clear variations in both the peak position and FWHM, as shown in Figs. 7(b) and 7(c), respectively. The peak position shifts by about 0.3 cm$^{-1}$ when the gate voltage is varied from 0.5 to –3.5 V, displaying a consistent and reproducible trend in this gate voltage range. This blue shift in the Raman band is accompanied by a slight peak broadening, with the FWHM increasing from ~ 7.8 to 8.4 cm$^{-1}$, corresponding to an increase of about 0.6 cm$^{-1}$. Although the Raman shift is relatively small (~0.3 cm$^{-1}$), it varies systematically with gate voltage. Despite the nominal spectral resolution of the Raman spectrometer is 0.5 cm$^{-1}$, Lorentzian fitting enabled determination of peak positions with a precision of ~ 0.1 cm$^{-1}$. Owing to the high signal-to-noise ratio of the spectra, the uncertainty in the extracted peak positions is less than 0.1 cm$^{-1}$. We note here that the electro-chemical doping of holes are confined to only a few nanometers near the surface, while the Raman signal is largely dominated by the bulk diamond. Consequently, the measured Raman response represents a volume-averaged effect.

Previous studies on boron-doped diamond films have reported redshifts and increased FWHM in the Raman spectra with increasing dopant concentration [21]. At high boron levels (>10$^{20}$ cm$^{-1}$), these changes are often accompanied by Fano interference [19]. In contrast, this study is the first, to our knowledge, to observe a blue shift in diamond Raman peaks with electrochemical hole doping. Similar blue / red shifts have been reported in graphene, phosphorene, MoS$_2$, and other 2D materials under doping conditions [15,17,18,31,32]. In particular, for electrochemically gated graphene transistors, the G-band stiffens due to the removal of the Kohn anomaly in the phonon dispersion at the Γ point, while the linewidth narrows because phonon decay channels are suppressed by the Pauli exclusion principle as electron or hole doping shifts the Fermi level into the conduction or valence band [15]. In the case of a top-gated phosphorene transistor, Chakraborty et al. [17] reported a pronounced electron–hole asymmetry in electron–phonon coupling. Specifically, they observed phonon softening along with an increase in the FWHM of the $A_g$ modes under electron doping, whereas hole doping produced no significant change in either the phonon frequencies or the FWHM. In contrast, the $B_{2g}$ mode remained largely unaffected in terms of both frequency and linewidth under both electron and hole doping conditions [17]. Similarly, an in-situ Raman study of a monolayer MoS$_2$ top-gated transistor reported phonon softening and linewidth broadening of the $A_{1g}$ mode under electron doping, while the other Raman-active $E_{2g}^1$ mode remained essentially unaffected [18]. Moreover, an in-situ Raman study on few-layer MoTe$_2$ top-gated transistors reported phonon softening and linewidth broadening of the $E_{2g}^1$ and $B_{2g}$ modes under hole doping, whereas the $A_1g$ mode exhibited slight phonon hardening accompanied by linewidth narrowing [31]. In addition, phonon stiffening in WSe$_2$ films was observed as a function of hole doping achieved through controlled variation of octadecyltrichlorosilane concentration on the surface [32].

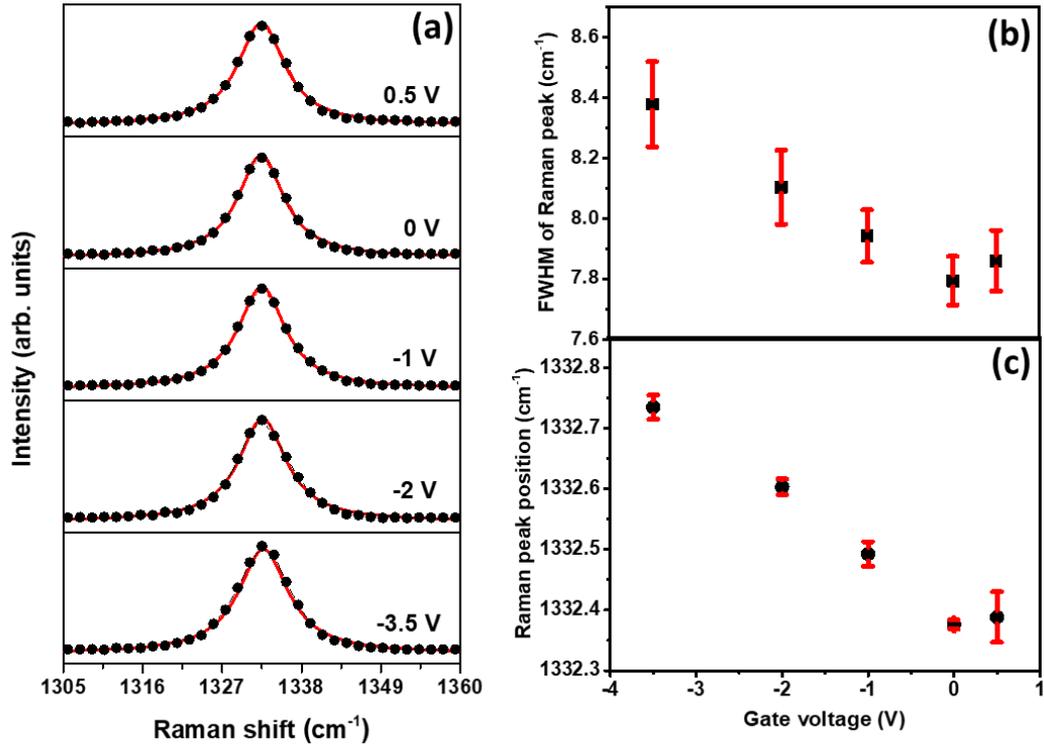

Fig. 7. (a) Raman spectra of diamond measured as a function of gate voltage for the H-diamond EGFET. The black dots are the experimental data and the red lines are fitted with Lorentzian function, (b) FWHM and (c) peak position of diamond Raman peak (1332 cm$^{-1}$) with respect to gate voltage.

From the above discussion, it is evident that phonon modes may undergo softening or stiffening under electron or hole doping, depending on the nature and strength of electron–phonon interactions in the system. In the present case, the observed phonon blue shift under electrochemical hole doping in hydrogen-terminated diamond is analogous to phenomena reported in other doped systems, such as electrostatically gated graphene and carbon nanotubes, where doping induces Raman mode stiffening accompanied by linewidth reduction [15,33]. In our system, the blue shift correlates with enhanced hole accumulation at negative gate bias, providing independent spectroscopic evidence of surface charge modulation. The observed linewidth broadening is likely associated with an inhomogeneous distribution of holes across the diamond surface, arising from surface disorder and roughness [27,34–36]. As hole density increases, the non-uniformity in hole accumulation becomes more pronounced, leading to greater variation in local phonon energies and lifetimes. It is worth noting here that the observed changes in Raman parameters are minimal, as the Raman signal predominantly originates from the bulk diamond, while hole accumulation is confined to a few nanometers from the surface [10,37]. Consequently, gating-induced phonon renormalization is confined to the surface and sub-surface regions, producing only a marginal effect on the film's bulk volume and, hence, only a slight change in the Raman signal. Nevertheless, a comprehensive theoretical investigation is necessary to fully elucidate the mechanism underlying the observed phonon stiffening and linewidth broadening in electrochemically hole-doped H-diamond.

## 4. Conclusions

In summary, the high surface conductivity (SC) in microcrystalline diamond films that are grown by hot filament chemical vapour deposition is accomplished by performing in-situ post growth H-termination. The partial O- termination on H-diamond (HD) is helped to transform the surface from hydrophobic to hydrophilic nature. In addition, the sheet resistance of HD is determined to increase from 7.6 to 18.7 kΩ/□ with reduction in hole density from 10.5 to 4.8 x $10^{12}$ cm$^{-2}$ with partial O-termination. Electrolyte-gated field effect transistors (EGFETs) were successfully fabricated on both pristine and partially O-terminated HD films using LiClO$_4$ and polyethylene oxide as electrolyte. From the electrical transfer characteristics of the devices, the source-drain saturation current is found to be ~ 40 and 14 times higher than that of cut-off region current. However, the areal capacitance of partially O-terminated HD EGFET is higher as compared to pristine HD EGFET due to higher wettability. While the overall performance of pristine HD EGFET is slightly superior, the partially O-terminated HD EGFET devices offer advantages in applications where the surface wettability is a crucial factor, such as chemical and biosensors applications. Additionally, the effect of electrochemical gating on phonon behaviour in H-diamond was studied over a gate voltage range of 0.5 to –3.5 V. The results reveal a gating-induced blue shift and broadening of the diamond Raman band, attributed to the Pauli-blocking of the phonon decay channels and the inhomogeneous hole distribution on the diamond surface. This work underscores the significant influence of partial O-termination on the performance of HD EGFET devices and elucidates the effect of electrochemical gating on the phonon behavior of H-diamond.

**Acknowledgement**

One of the authors, N.M.S, acknowledges IGCAR, DAE, Government of India for research fellowship.

**Funding**

The authors declare that no funds, grants, or other support were received during the preparation of this manuscript.

**Author Contributions**

- N. Mohasin Sulthana – Investigation, Formal analysis, Writing – original draft,

- P.K. Ajikumar - Investigation, Writing – review and editing

- K. Ganesan – Conceptualization, methodology, Formal analysis, Supervision, Writing – original draft, Writing – review and editing